\documentstyle[12pt]{article}
\newcommand{\ncm}{\newcommand}
\ncm{\oH}{\bar{H}}
\ncm{\us}{\quad\mbox{using}\quad}
\ncm{\ra}{\rightarrow}
\ncm{\ot}{\otimes}
\ncm{\DH}{D(H)}
\ncm{\DW}{D^{\omega}(H)}
\ncm{\TH}{T(\oH)}
\ncm{\im}{\imath}
\ncm{\ba}{\begin{array}}
\ncm{\ea}{\end{array}}
\ncm{\ul}{\underline}
\ncm{\ol}{\overline}

\def\hoek{\hbox{\vrule height 2.5ex depth 0pt \vrule width 2.5ex height .4pt
 depth 0pt}}
\def\haak#1#2{
\mathop{\hoek\llap{\vbox to 2.5ex{ \vfil
\hbox{$\scriptstyle#1$\hskip 2.8ex} \vfil}}}
\limits_{#2} }
\def\hook#1#2{\setbox0=\hbox{$\scriptstyle#1$}
\hskip\wd0\haak{\box0}{#2}}
\ncm{\str}{\rule{0cm}{3.5mm}}
\ncm{\om}{\omega}
\ncm{\ep}{\epsilon}
\newlength{\extraspace}
\newlength{\extraspaces}
\newcommand{\be}{\begin{equation}
\addtolength{\abovedisplayskip}{\extraspaces}
\addtolength{\belowdisplayskip}{\extraspaces}
\addtolength{\abovedisplayshortskip}{\extraspace}
\addtolength{\belowdisplayshortskip}{\extraspace}}
\newcommand{\ee}{\end{equation}}

\newcommand{\bea}{\begin{eqnarray}
\addtolength{\abovedisplayskip}{\extraspaces}
\addtolength{\belowdisplayskip}{\extraspaces}
\addtolength{\abovedisplayshortskip}{\extraspace}
\addtolength{\belowdisplayshortskip}{\extraspace}}
\newcommand{\eea}{\end{eqnarray}}
\newcommand{\beas}{\begin{eqnarray*}
\addtolength{\abovedisplayskip}{\extraspaces}
\addtolength{\belowdisplayskip}{\extraspaces}
\addtolength{\abovedisplayshortskip}{\extraspace}
\addtolength{\belowdisplayshortskip}{\extraspace}}
\newcommand{\eeas}{\end{eqnarray*}}

\ncm{\al}{\alpha}
\ncm{\bt}{\beta}
\ncm{\gm}{\gamma}
\ncm{\dl}{\delta}
\ncm{\varep}{\varepsilon}
\ncm{\zt}{\zeta}
\ncm{\et}{\eta}
\ncm{\th}{\theta}
\ncm{\kp}{\kappa}
\ncm{\lm}{\lambda}
\ncm{\rh}{\rho}
\ncm{\hl}{\hline}
\ncm{\sg}{\sigma}
\ncm{\ta}{\tau}
\ncm{\ph}{\phi}
\ncm{\phv}{\varphi}
\ncm{\ch}{\chi}
\ncm{\ps}{\psi}
\ncm{\nn}{\nonumber}
\title{Anyons in discrete gauge theories\\ with Chern-Simons terms}
\author{F.Alexander Bais\thanks{email: bais@phys.uva.nl} ,
Peter van Driel\thanks{email: vandriel@phys.uva.nl}
$\,$ and
Mark de Wild Propitius\thanks{email: mdwp@phys.uva.nl}
\\ Instituut voor Theoretische Fysica\\Valckenierstraat 65\\
1018XE Amsterdam}
\date{February 1992}
\begin{document}
\maketitle
\begin{abstract}
We study the effect of a Chern-Simons term in a theory with discrete gauge
group $H$, which in (2+1)-dimensional space time describes (non-abelian)
anyons. As in a previous paper \cite{bpm1}, we emphasize the underlying
algebraic structure, namely the Hopf algebra $D(H)$. We argue on physical
grounds that the addition of a Chern-Simons term in the action leads to a
non-trivial 3-cocycle on $D(H)$. Accordingly, the physically inequivalent
models are
labelled by the elements of the cohomology group  $H^3(H, U(1))$. It depends
periodically on the coefficient of the Chern-Simons term which model is
realized. This establishes  a relation with
the discrete topological field theories of Dijkgraaf and Witten. Some
representative examples are worked out explicitly.
\end{abstract}
\vspace*{2cm}
Preprint ITFA-92-8 submitted to: Physics Letters B
\newpage  \noindent

\section{Introduction}
It is  well established by now that quantumstatistics for identical particles
in (2+1)-dimensional space time can be quite exotic. The reason is that
interchanges of identical particles in the plane are not organized by the
symmetric group, but rather by the braid group~\cite{wu}.
The 1-dimensional unitary irreducible
representations of the braid group are labelled by an angular parameter
$\Theta\in[0,2\pi)$. Interchanging two identical particles in
such a representation yields a phase $\exp \im \Theta$, and we see
that in the plane we have the possibility of quantumstatistics
intermediate~\cite{lein}
between the familiar examples bosons $\Theta=0$ and fermions $\Theta=\pi$.
Particles obeying intermediate quantumstatistics have been named
anyons~\cite{anyon}.
Of course, particles with quantumstatistical properties
associated with higher dimensional unitary irreducible representations of the
braid group are also conceivable. We call these
non-abelian anyons.

Anyons and non-abelian anyons may appear as topological excitations in
theories where a continuous gauge group $G$ is spontaneously broken down to
a discrete group $H$. Their long range topological interactions are
effectively described by a discrete $H$ gauge theory~\cite{amw}.
The underlying algebraic structure of these theories strongly resembles
that of conformal field theories, in particular holomorphic
orbifold models~\cite{dvvv}. In a previous paper~\cite{bpm1},
we showed that this underlying structure is
the quasi-triangular Hopf algebra $D(H)$ \cite{dpr1}.
This insight allowed us to give a
complete description of the invariant couplings (i.e.\ fusion rules)
and the two particle Aharonov-Bohm scattering (i.e.\ braiding properties)
of these anyonic excitations.
In the present paper we include
a Chern-Simons term and analyse its physical consequences.
What we will show -~and
work out explicitly for  $H \simeq Z_N$~- is
that the Chern-Simons term leads to a deformation of the
algebra $D(H)$ by a non-trivial 3-cocycle. This implies
among others, that the dependence of the effective theory
on the quantized Chern-Simons parameter $\mu$ is periodic. This
provides a correspondence with the discrete topological field
theories introduced by Dijkgraaf and Witten~\cite{diwi}.

The paper is organized as follows. In section~2, we  briefly introduce a
$G\simeq SU(2)$ gauge model with Chern-Simons term.
Next we determine from purely physical arguments the fusion- and
braiding properties of the superselection sectors for $H\simeq Z_N$.
The resulting mathematical structure is precisely the quasi-Hopf
algebra $\DW$, with $\omega$  a 3-cocycle. This structure is
discussed in section~3. In section~4 we treat some examples in detail, starting
with $D^\omega(Z_N)$. We show that this algebra indeed gives the same fusion-
and braiding properties as established in section~2 for a $Z_N$ gauge theory
with Chern-Simons term. We also consider an  example describing non-abelian
anyons, namely
$H\simeq \bar{D}_2$. The paper closes with some concluding remarks.

\section{The physical model}           \label{model}

Although our analysis holds in general,
we restrict our considerations to a model with
gauge group $G \simeq SU(2)$ to make the discussion explicit.
The starting point is the lagrangian
\be  \label{action}
{\cal L}= -\frac{1}{4}F_{\rho\nu}^a F^{a\,\rho\nu} + (D_\rho \Phi)^*
\cdot(D^\rho \Phi) - V(\Phi) + \frac{\mu}{4} \epsilon^{\kappa\sigma\rho}
[F_{\kappa\sigma}^a A^a_\rho + \frac{2}{3}e\epsilon^{abc}A_\kappa^a
A_\sigma^b A_\rho^c ].
\ee
We are working in (2+1)-dimensional space time, so greek indices
run from 0 to 2. Latin indices  label the three (hermitean)
generators of  $G \simeq SU(2)$.  The covariant
derivative takes the form $D_\rho \Phi = (\partial_\rho +ieA_\rho^a T^a)\Phi$,
with the generators $T^a$ of $SU(2)$  in the representation of the Higgs field
$\Phi$. The last term in~(\ref{action}) is the Chern-Simons term with
$\epsilon$  the completely anti-symmetric three dimensional
Levi-Civita tensor. Requiring this theory to be  gauge invariant at the
quantum level, implies a quantization condition~\cite{des1} for the
topological mass  $\mu$
\be \label{mu}
\mu = pe^2/4\pi \;\;\;\;\;\; \mbox{with} \;\;\; p\in Z.
\ee
We set out to study the effect of the Chern-Simons term on the excitations
in the discrete $H$ gauge theory obtained after breaking
$G \simeq SU(2)$ down to the finite group $H$.
It is illustrative to do this for a simple example, namely the discrete $Z_N$
gauge theory that arises from the simple symmetry breaking scheme
$SU(2)\rightarrow U(1)\rightarrow Z_N$. Such a scheme can be realized by a
proper choice of the Higgs field $\Phi$ and its potential $V(\Phi)$.

In the low energy $U(1)$ regime, the theory is governed by the effective
lagrangian
\be                          \label{effaction}
{\cal L}_{ef\!f}= -\frac{1}{4}F_{\rho\nu} F^{\rho\nu} + (D_\rho
\Phi)^*(D^\rho \Phi) - V(\Phi) + \frac{\mu}{4} \epsilon^{\kappa\sigma\rho}
F_{\kappa\sigma} A_{\rho}.
\ee
We have omitted the massive modes in this strictly abelian model,
and more important the instantons labelled by $\pi_2(SU(2)/U(1))\simeq Z$.
These instantons will play a profound role in the determination of the
topological superselection sectors of the theory later on.
Variation with respect to the vectorpotential
$A^\sigma$, yields the  field equation
\be                      \label{fieldequation}
\partial_\rho F^{\rho\sigma} +
\frac{\mu}{2}\epsilon^{\sigma\tau\rho}F_{\tau\rho} = j^\sigma.
\ee
We want to consider all sectors of the theory.
So from here on, we will take $j^\sigma$ to
stand not just for the conserved current associated with the Higgs field
$\Phi$, but also for contributions of possible other $SU(2)$  matter fields,
which for notational convenience were not displayed
in~(\ref{action}).
The different currents occuring in~(\ref{fieldequation})
are separately
conserved. Integrating the zeroth component of~(\ref{fieldequation}) over two
dimensional space, leads to the following relation between the corresponding
conserved charges
\be                     \label{Q}
Q =q+ \mu \phi,
\ee
with $Q=\int\! d^2x\,\vec{\nabla}\!\cdot\!\vec{E}$ the electric charge,
$q=\int\!
d^2x \, j^0 $ the global $U(1)$ charge, and their difference $q_{ind}=\mu
\phi$ the charge induced by the Chern-Simons term. We denote the total
magnetic flux by $\phi\equiv \int \! d^2x \,\epsilon^{ij}\partial_i A^j$.

In the $U(1)$ phase electric charges are quantized in units $e/2$.
Although there are no stable magnetic fluxes
in this phase  ($\pi_1(SU(2)/U(1))\simeq 0$), there are instantons labelled by
$\pi_2(SU(2)/U(1)) \simeq \pi_2(S^2)\simeq Z$.
In  3 euclidean dimensions these instantons are monopoles carrying  magnetic
charge $g=4\pi k/e$ with $k\in Z$. Now suppose that in a next
stage of symmetry breaking  a component of the the Higgs field $\Phi$  which
carries a $U(1)$ charge $Ne/2$ condenses. This gives rise to the subsequent
symmetry breakdown  $U(1)\rightarrow Z_N$.
In this $Z_N$ phase we have  magnetic flux excitations
$\phi= 4\pi m/Ne$ with $m \in \pi_1(U(1)/Z_N)\simeq Z$.
In the present context this is not the whole story. The proper labelling of
the magnetic flux sectors in the full theory is by $\pi_1(SU(2)/Z_N)\simeq
Z_N$.
The apparent difference can be
understood if the role of the instantons in the model is taken into account.
In the $Z_N$ phase these configurations describe tunneling
events where magnetic flux excitations  $\phi=4\pi/e$ decay into the vacuum.
So in the $SU(2)\rightarrow U(1)\rightarrow Z_N$ model where instantons are
mandatory, flux is conserved modulo $N$. In a strict $G\simeq U(1)\rightarrow
Z_N$ model, the flux is conserved (i.e.\ $m\in Z$) if we do not put an
instanton in the theory, or conserved modulo $N$ if we do put an instanton in
by hand. As we shall see later this choice also has an important bearing
on the question whether the Chern-Simons parameter $\mu$ has to
be quantized in a $G\simeq U(1)$ theory.

Let us now turn to the fate of the electric charges in the $Z_N$ phase.
As has been argued in~\cite{amw}, there are charges surviving the screening
mechanism accompanying the breakdown of the continuous $U(1)$ symmetry.
The point is, the condensate of $Ne/2$ charges
can only completely screen charges that are multiples of $Ne/2$. So the
different $Z_N$ charges are $q=ne/2$ with $n\in Z$ modulo N.
Since the gauge fields are massive, these charges do {\em not} carry
long range Coulomb fields, and are unobservable classically. However,
by means of
Aharonov-Bohm scattering processes~\cite{ver1} with the magnetic fluxes,
 we are able to detect the $Z_N$ charges quantummechanically. The
essential quantity featuring in the cross-sections for these  scattering
processes is the quantum phase picked up by such a charge after encircling
a magnetic flux once. In the absence of a Chern-Simons term, this quantum phase
generated by the coupling $-j^\rho A_\rho$, takes the value
\be        \label{phase}
e^{\im (q_1 \phi_2 + q_2 \phi_1)} = e^{\im \frac{2\pi}{N}
(n_1m_2+n_2m_1)}.
\ee
Here we considered the process of carrying a flux-charge composite
$(m_1,n_1) \equiv (\phi_1 \!=\! 4\pi m_1/Ne,\, q_1 \!=\! n_1 e/2)$
around a composite $(m_2,n_2)$.  The phase~(\ref{phase}) indicating  that
these dyonic composites behave as anyons, can be interpreted
as the product of a $Z_N$ gauge transformation on the charge $n_1$ (mod $N$)
by an element $m_2$ and a transformation on the charge $n_2$ (mod $N$) by an
element $m_1$. In our model, the Aharonov-Bohm
scattering processes are also the only way to detect the magnetic
fluxes. Using~(\ref{phase}), we  therefore conclude that for
$\mu=0$ the flux $m$ is defined modulo  $N$ as well.
For a $Z_4$ gauge theory,
for instance, these observations lead us to the spectrum depicted in
the dashed box in Figure~1(a), where in this case the box has strictly
periodic boundary conditions.

In the presence of a Chern-Simons term, the situation is slightly more
involved. Instead of the ordinary coupling $-j^\rho A_\rho$, we find
in~(\ref{effaction}) the coupling
\be \label{coupling}
{\cal L}_{int}= -(j^\rho-\frac{\mu}{4}\epsilon^{\rho\kappa\sigma}
F_{\kappa\sigma})A_\rho.
\ee
This means that instead of~(\ref{phase}), the same process generates
the phase
\be    \label{csphase}
e^{\im (\tilde{Q}_1\phi_2 + \tilde{Q}_2 \phi_1)} =
e^{\im \frac{2\pi}{N}((n_1+\frac{p}{N}m_1)m_2 +
(n_2+\frac{p}{N}m_2)m_1)},
\ee
where we introduced the charge
\be    \label{noether}
\tilde{Q} \equiv q + \frac{\mu}{2} \phi.
\ee
If we now, as before, interpret the total phase (\ref{csphase}) as two gauge
transformations, we see that the generator of the $Z_N$ transformations is the
shifted charge $\tilde{Q}$. We will  verify this later on by
computing the Noether charge for the gauge transformations.
We  now argue that~(\ref{csphase}) allows for the conclusion
that the spectrum of a
$Z_N$ gauge theory in the presence of a Chern-Simons term can also be
reduced to the box depicted in Figure~1(a), although the boundary condition
in the $\phi$ direction is twisted.
Consider the process in which we encircle a composite
$(m_1+m_2,n_1+n_2)$ by an arbitrary composite $(m_3,n_3)$. The sum $m_1+m_2$
does not necessarily lay between $0$ and $N-1$. Using the notation $[m_1+m_2]$
for $(m_1+m_2)\,\,\,\mbox{modulo}\,\, N $, chosen between $0$ and $N-1$,
we can rewrite the phase~(\ref{csphase})  as
\be \label{modulo}
e^{ \im (\tilde{Q}_3\phi_1 + \tilde{Q}_1 \phi_3 + \tilde{Q}_3\phi_2
+ \tilde{Q}_2 \phi_3 )} = e^{ \im (\tilde{Q}_3\phi_{12} +
\tilde{Q}_{12}\phi_3)},
\ee
with the definitions
\bea
\tilde{Q}_{12} &\equiv& q_{12} + \frac{\mu}{2} \phi_{12} \label{fusQt}\\
   & \equiv &
([n_1+n_2+\frac{2p}{N}(m_1+m_2-[m_1+m_2])] + \frac{p}{N}[m_1+m_2])
\;\frac{e}{2}
 \nn \\
\phi_{12} &\equiv& \frac{4\pi}{Ne}[m_1+m_2].   \label{fusphi}
\eea
The equations~(\ref{fusQt}) and~(\ref{fusphi}) express the way charges
and fluxes `add', i.e.\ they specify the {\em fusion rules}
for the excitations in our model. In terms of the quantumstates
$|m,n\!>$, these read
\be \label{csfusion}
|m_1,n_1\!>\!\times|m_2,n_2\!>\,=
|[m_1+m_2],[n_1+n_2+\frac{2p}{N}(m_1+m_2-[m_1+m_2])]>.
\ee
We conclude that the spectrum of a $Z_N$ gauge theory in the presence
of a Chern-Simons term with $\mu$ quantized as~(\ref{mu}), again boils
down to the
excitations in the dashed box in Figure~1(a). If $p\neq 0$ however, the modulo
calculus to map excitations inside the box is not simply straight
along the $\phi$ and $q$ axes. For example,
the excitation $(4\pi/e,0)$ obtained by fusing two
fluxes $(2\pi/e,0)$ is identified with  $(0,e)$. In other words, the
periodicity in the $\phi$ direction is twisted by $2p$ units, implying that if
$p\neq 0$ the charge $q$ is no longer strictly conserved modulo $N$.
The same is true for the  charge $\tilde{Q}$, which is
twisted by $p$ units as is clear from Figure~1(b). It is in fact the electric
charge $Q$ defined in~(\ref{Q}), which stands out in this respect. As
Figure~1(c) indicates  $\phi$ and $Q$ are independently conserved modulo $N$.
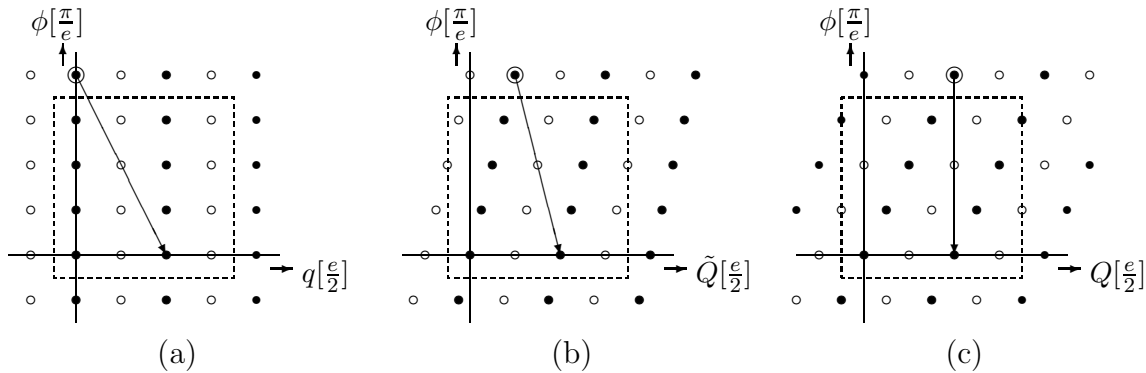
\begin{figure}[t]
\begin{center}
\begin{picture}(85,80)(-15,-15)
\put(-5,-5){\dashbox(40,40)[t]{}}
\put(-15,0){\line(1,0){60}}
\put(0,-15){\line(0,1){60}}
\thinlines
\multiput(-10,-10)(0,10){6}{\multiput(0,0)(10,0){5}{\circle{1.6}}}
\multiput(0,-10)(0,10){6}{\multiput(0,0)(20,0){3}{\circle*{2.0}}}
\put(0,40){\circle{3.6}}
\put(0,40){\vector(1,-2){20}}
\put(-3,42){\vector(0,1){5}}
\put(43,-3){\vector(1,0){5}}
\put(-10,50){\small$\phi[\frac{\pi}{e}]$}
\put(50,-6){\small$q[\frac{e}{2}]$}
\put(18,-24){(a)}
\end{picture}
\begin{picture}(85,80)(-15,-15)
\put(-5,-5){\dashbox(40,40)[t]{}}
\put(-15,0){\line(1,0){60}}
\put(0,-15){\line(0,1){60}}
\thinlines
\multiput(-12.5,-10)(2.5,10){6}{\multiput(0,0)(10,0){6}{\circle{1.6}}}
\multiput(-2.5,-10)(2.5,10){6}{\multiput(0,0)(20,0){3}{\circle*{2.0}}}
\put(-3,42){\vector(0,1){5}}
\put(43,-3){\vector(1,0){5}}
\put(10,40){\circle{3.6}}
\put(10,40){\vector(1,-4){10}}
\put(-10,50){\small$\phi[\frac{\pi}{e}]$}
\put(50,-6){\small$\tilde{Q}[\frac{e}{2}]$}
\put(18,-24){(b)}
\end{picture}
\begin{picture}(85,80)(-15,-15)
\put(-5,-5){\dashbox(40,40)[t]{}}
\put(-15,0){\line(1,0){60}}
\put(0,-15){\line(0,1){60}}
\thinlines
\multiput(-15,-10)(5,10){6}{\multiput(0,0)(10,0){5}{\circle{1.6}}}
\multiput(-5,-10)(5,10){6}{\multiput(0,0)(20,0){2}{\circle*{2.0}}}
\multiput(-15,10)(5,10){4}{\multiput(0,0)(20,0){1}{\circle*{2.0}}}
\multiput(35,-10)(5,10){4}{\multiput(0,0)(20,0){1}{\circle*{2.0}}}
\put(-3,42){\vector(0,1){5}}
\put(43,-3){\vector(1,0){5}}
\put(20,40){\circle{3.6}}
\put(20,40){\vector(0,-1){40}}
\put(-10,50){\small$\phi[\frac{\pi}{e}]$}
\put(50,-6){\small$Q[\frac{e}{2}]$}
\put(18,-24){(c)}
\end{picture}
\vspace{0.5cm}
\caption{\em The spectrum of a $Z_4$ gauge theory. We depict the flux $\phi$
against the global $U(1)$ charge $q$, the  Noether charge $\tilde{Q}$
and the electric charge $Q$ respectively. The
Chern-Simons parameter $\mu$ is set to its minimal non-trivial value
$\mu=e^2/4\pi$, i.e.\ $p=1$. The identification of the encircled excitation
with an excitation inside the dashed box is indicated with an arrow.}
\end{center}
\end{figure}

Note that the  fusion rules~(\ref{csfusion}) are periodic in $p$ with period
$N/2$ for $N$ even. The  braid properties~(\ref{csphase}) of the spectrum in
contrast, are periodic in $p$ with period $N$. {\em Thus there are only $N$
different $Z_N$ Chern-Simons gauge theories if $\mu$ is quantized
as in~(\ref{mu}).}

At this point it is worthwile to make a little digression and return
to a question we alluded to before, namely whether the parameter $\mu$ should
be quantized or not in a $G\simeq U(1)\rightarrow Z_N$ theory.
We can distinguish a few possibilities. Let us first assume that the theory
contains instantons. Recall that these are the monopoles in 3 euclidean
dimensions.
Hence following Dirac's argument~\cite{dirac}, this implies that the charges
$q$ are quantized, and as we argued before magnetic flux is conserved
modulo $N$.
If we now in addition assume that the Chern-Simons parameter $p$ is not
quantized (i.e.\ $p \not\in Z$),  we run into a contradiction.
As follows from Figure~1(a), a tunneling event by means of an instanton
involving the
magnetic flux $\phi=4\pi/e$ would give a charge $2p$. This is inconsistent
with the earlier observation that only (half) integral charges were present.
{\em The conclusion is that if instantons are present, then $\mu$ has to be
quantized accordingly.} Interestingly enough the (2+1)-dimensional version
of Dirac's quantization condition gives the correct $\mu$ quantization
for general gauge groups.
We now turn to the  case where instantons are absent.
The simplest situation arises if the charges $q$ are not quantized. Then
flux is a $Z$-quantumnumber and $p$ need not be quantized. Finally, if
charges are quantized $q=ne/2$ with $n\in Z$ and
$p$ is irrational, then all fluxes $m\in Z$ are distinguishable.
If $p=k/l$ with $k$ and $l$ relative primes however, then we find
$m=m \; \mbox{modulo}\; lN/\gcd(k,N)$, where
$\gcd(k,N)$ denotes the greatest common divisor of $k$ and $N$.

We conclude this section with a direct computation of the Noether charge
generating the residual $Z_N$ transformation in the presence of a Chern-Simons
term. First we observe that the lagrangian~(\ref{effaction}) is not invariant
under a gauge transformation
\beas
\Phi(x) &\longmapsto& e^{\im \frac{Ne}{2}\Omega(x)} \; \Phi(x) \\
A_\rho (x) &\longmapsto& A_\rho (x)-\partial_\rho \Omega (x),
\eeas
but rather changes by a total derivative
\[
\delta_{\Omega}{\cal L}_{ef\!f}=-\frac{\mu}{4}\partial_{\sigma}
(\epsilon^{\sigma\rho\tau} F_{\rho\tau}\Omega).
\]
This term should be substracted from the usual Noether current
\beas
\tilde{\jmath}^\sigma_{\Omega} &=&
-\frac{\partial{\cal L}_{ef\!f}}{\partial(\partial_\sigma \Phi)} \delta\Phi
-\frac{\partial{\cal L}_{ef\!f}}{\partial(\partial_\sigma \Phi^\dagger)}
\delta\Phi^\dagger
-\frac{\partial{\cal L}_{ef\!f}}{\partial(\partial_\sigma A_\rho)}
\delta A_\rho +\frac{\mu}{4}\epsilon^{\sigma\rho\tau}F_{\rho\tau}\Omega \\
 &=& (j^\sigma-\frac{\mu}{4}\epsilon^{\sigma\rho\tau}F_{\rho\tau}) \Omega
+(-F^{\sigma\rho} + \frac{\mu}{2}\epsilon^{\sigma\rho\tau}A_\tau)\partial_\rho
\Omega.
\eeas
The corresponding Noether charge can be calculated either
by partially integrating and subsequently using~(\ref{fieldequation}), or
alternatively by taking $\Omega$ constant.
We obtain the form $\int \! d^2 x \, \tilde{\jmath}\,^0_{\Omega} =\Omega
(\infty) \tilde{Q}$, with $\tilde{Q}$ defined in~(\ref{noether}). From this
exercise we learn that a gauge transformation $\Omega(\infty)= 4\pi l/Ne$
(with $l=0,\ldots,N-1$) corresponding to the element $l$ of the residual
symmetry group $Z_N$, acts on the quantumstate $|m,n\!>$ as
\be \label{ga+co}
U(l) \,|m,n\!> \;=\;e^{\im \frac{4\pi}{Ne}l\tilde{Q}}\,|m,n\!>\;=\;
e^{\im \frac{2\pi}{N}l(n+p\frac{m}{N})} \, |m,n\!>.
\ee
We  see that the charge $\tilde{Q}$ that generates the residual gauge
transformation~(\ref{ga+co}) is indeed the same as the charge that enters the
Aharonov-Bohm phase~(\ref{csphase}). The fact that the Noether charge
$\tilde{Q}$, rather than the electric charge $Q$, features in the
Aharonov-Bohm phase is well known. See for example~\cite{wil4}.

The action~(\ref{ga+co}) of the $Z_N$ gauge group  in the presence of
a Chern-Simons term is rather unusual:
\begin{itemize}
\item
The fractional spectrum of $\tilde{Q}$ indicates
that the residual gauge
transformations no longer form an ordinary $Z_N$ representation. Indeed,
performing two successive gauge transformations yields
\be         \label{ray}
U(k) \cdot U(l) = e^{\im \frac{4\pi}{Ne}(k+l - [k+l])\tilde{Q}}\;
U([l+k]) =e^{\im \frac{2\pi p}{N^2}m(k+l-[k+l])}\;U([l+k]).
\ee
Here we used the form that $\tilde{Q}$ takes for the excitations in our model,
as we did in~(\ref{ga+co}). It is easily verified that the additional phase
in~(\ref{ray}) is a 2-cocycle~\cite{jack}, so associativity is preserved.
We conclude that the residual gauge
transformations $U(l)$ constitute a projective- or ray representation
of $Z_N$ in the presence of a Chern-Simons term. These ray representations are
trivial though;  the extra phases can be eliminated by redefining the $U$'s.
\item
A similar observation can be made if we implement a residual gauge
transformation~(\ref{ga+co}) on a two-particle state
\bea \label{twopart}
U(l) (|m_1,n_1\!>\!|m_2,n_2\!>\!)\!\!
&=& \!\!
e^{\im \frac{4\pi}{Ne}l \tilde{Q}_{12}} (|m_1,n_1\!>\!|m_2,n_2\!>\!)\\
\!\!&=& \!\!
\nonumber
e^{\im \frac{2\pi p}{N^2}l(m_1+m_2-[m_1+m_2])} (\!U(l)|m_1,
n_1\!>\!U(l)|m_2,n_2\!>\!)                                 .
\eea
This is obviously not the same as the ordinary product of the
gauge transformation on the individual states $|m_1,n_1\!\!>$ and
$|m_2,n_2\!\!>$. Stated mathematically: the Chern-Simons term
also alters the {\em co-multiplication} by a 2-cocycle. These extra
phases could have been eliminated  by redefining the $U$'s as well.
It is not possible however, to make the phases in~(\ref{ray})
and~(\ref{twopart}) disappear simultaneously.
\item
There are now two isomorphic ways to describe three-particle states, namely\\
$(|m_1,n_1\!>\!\!(|m_2,n_2\!>\!\!|m_3,n_3\!>))$ and
$((|m_1,n_1\!>\!\!|m_2,n_2\!>)|m_3,n_3\!>)$.
The  isomorphism (expressing {\em quasi}-coassociativity)
involves a 3-cocycle $\omega$.
The 2-cocycles appearing in~(\ref{ray})
and~(\ref{twopart}) stem from this 3-cocycle. All this will be argued
in greater detail in the following sections.
\end{itemize}
The  algebraic structure underlying a discrete $H$ gauge theory without a
Chern-Simons term~\cite{bpm1} is the Hopf algebra $D(H)$.
In the following sections, we will show that for $H \simeq Z_N$
the inclusion of a Chern-Simons term introduces the just mentioned
non-trivial 3-cocycle $\omega$ on $D(Z_N)$. The generalisation of this
procedure to $D(H)$ with non-abelian $H$, and thus the effect of a Chern-Simons
term for non-abelian fluxes as introduced in~\cite{bais}, is straightforward.

\section{The quasi-Hopf algebra $\DW$}   \label{Hopf}
The distinguishing features of the quasi-Hopf algebra $\DW$ are {\em
quasi-coassociativity} and {\em quasitriangularity}. Quasi-coassociativity
means that there exists an invertible element $\varphi\in
D^{\omega}(H)^{\otimes 3}$, such that
\be \label{quasicoas}
(id \ot \Delta)\Delta(a)=\varphi\cdot (\Delta \ot id)\Delta(a)
\cdot\varphi^{-1}\;\;\;\;\;\;\forall a\in\DW.
\ee
If $(\Pi_i, V_i)$ denote representations of $\DW$, then~(\ref{quasicoas})
establishes the equivalence of the
representations  $\Pi_1 \otimes (\Pi_2 \otimes \Pi_3)$ and
$(\Pi_1 \otimes \Pi_2) \otimes \Pi_3$ by means of the non-trivial isomorphism
\bea                              \label{fi}
\Phi :\qquad (V_1 \ot V_2) \ot V_3 & \longrightarrow & V_1 \ot (V_2 \ot V_3) \\
\qquad\qquad\qquad  v_1 \ot v_2 \ot v_3 & \longmapsto &
\Pi_1 \! \ot \! \Pi_2 \! \ot \Pi_3 \,(\varphi)\;v_1 \ot v_2 \ot v_3.
\nonumber
\eea
The fact that $\DW$ is quasitriangular, implies something similar.
It stands for the existence of an element $R\in\DW \ot \DW$, that satisfies
among others
\be
\Delta'(a)  = R\cdot \Delta(a)\cdot R^{-1}\;\;\;\;\;\;\forall a\in\DW,
\label{Rmat}
\ee
where we defined $\Delta'(a) \equiv \sum_i b_i \ot a_i$, if
$\Delta(a)= \sum_i a_i \ot b_i$.
At the level of representations, relation~(\ref{Rmat})
reflects the equivalence of $\Pi_1 \otimes \Pi_2$ and $\Pi_2 \otimes \Pi_1$.
This equivalence is obtained through
the non-trivial isomorphism $\Pi_1 \otimes \Pi_2 (R)$ from  $V_1 \otimes V_2$
into $V_1 \otimes V_2$, followed by the simple permutation
\bea \label{permutation}
\sigma :\qquad V_1 \otimes V_2&\longrightarrow& V_2 \otimes V_1.
\eea
We now turn to the explicit realization of these abstract notions in
the algebra $\DW$. This algebra is spanned by the
basis $\{\hook{g}{x}\}_{g,x\in H}$.
In terms of these basis elements the multiplication and
co-multiplication are given by
\be \label{algebra}
\hook{g}{x}.\hook{h}{y} =\delta_{g,xhx^{-1}}\hook{g}{xy} \theta_g(x,y)\; ,
\qquad\qquad
\Delta(\,\hook{g}{x})=\sum_{\{h,k|hk=g\}}\hook{h}{x}\otimes\hook{k}{x}
\gamma_x(h,k) \; ,
\ee
where the $U(1)$ valued functions $\theta_g$ and $\gamma_g$ equal $1$ if $g$
or one of their variables is the unit $e$ of $H$.
The elements $\varphi$ and $R$ take the form
\be                    \label{Rmatrix}
\varphi=\sum_{g,h,k}\,\omega^{-1}(g,h,k)\,
\hook{g}{e}\otimes\hook{h}{e}\otimes\hook{k}{e}\; , \;\;\;\;\;\;\;\;\;
R=\sum_{g,h}\,\hook{g}{e}\otimes\hook{h}{g} \; ,
\ee
where $\omega$ is again $U(1)$ valued. Consistency of the use of $R$ and
$\varphi$ in arbitrary tensor products implies \cite{dpr1}
\bea
\label{theta}
\theta_g(x,y) &=& \frac{\omega(g,x,y)\;\omega(x,y,(xy)^{-1}gxy)}
{\omega(x,x^{-1}gx,y)}  \\
\gamma_x(g,h)&=&\frac{\omega(g,h,x)\;\omega(x,x^{-1}gx,x^{-1}hx)}{\omega(g,x,
x^{-1}hx)}\; , \label{gamma}
\eea
where $\omega$ is a 3-cocycle
\be
\label{pentagon}
\omega(g,h,k)\;\omega(g,hk,l)\;\omega(h,k,l)=\omega(gh,k,l)\;\omega(g,h,kl).
\ee
Equation (\ref{pentagon}) determines $\omega$ uniquely up to the
coboundary $\delta\beta$
\be
\label{twist}
\omega(g,h,k)\;\;\longmapsto\;\;\frac{\beta(g,h)\;\beta(gh,k)}
{\beta(h,k)\;\beta(g,hk)}\;\omega(g,h,k).
\ee
Equivalence classes of solutions of (\ref{pentagon}), and hence the different
algebras $\DW$, are labelled by the third
cohomology group $H^3(H,U(1))$. The equivalence of $\DW$ and
$D^{\omega\delta\beta}(H)$ is argued as follows: If we accompany the
transformation~(\ref{twist}) on $\DW$ by the basis transformation
\be
\label{basistr}
\hook{g}{x} \;\;\longmapsto \;\;\frac{\beta(x,x^{-1}gx)}{\beta(g,x)}
\;\;\;\hook{g}{x} \; ,
\ee
we can readily verify that the algebra $D^{\omega\delta\beta}(H)$ is
obtained from $\DW$ by a twist with the element
$F=\sum_{g,h} \beta^{-1}(g,h)\, \hook{g}{e} \ot \hook{h}{e}$
\bea \label{F}
\Delta(\,\hook{g}{x}) &\longmapsto& F \cdot \Delta(\,\hook{g}{x}) \cdot F^{-1}
\\
R &\longmapsto& F_{21} \cdot R \cdot F^{-1} \nonumber \\
\varphi &\longmapsto& ({\bf 1}\ot F) \cdot (id \ot \Delta)\,(F) \cdot
\varphi \cdot
(\Delta \ot id)\,(F^{-1}) \cdot (F^{-1} \ot {\bf 1}) \;, \nonumber
\eea
with $F_{21} \equiv \sum_{g,h}\, \beta^{-1}(g,h) \hook{h}{e} \ot \hook{g}{e}\,
$. We will show shortly that this implies that the representation theory for
the algebra $\DW$ is  equivalent to that for
$D^{\omega\delta\beta}(H)$.  There is therefore no way to
distinguish these two algebras.
It is noteworthy that you
can always choose a twist~(\ref{twist}), such that $\theta$~(\ref{theta}) and
$\gamma$~(\ref{gamma}) indeed equal $1$ whenever one of the entries is the
unit $e$ of $H$.\footnote{Of course, you can choose not to fix
this `gauge freedom'. In this case, the construction of the quasi-Hopf algebra
$\DW$ still goes along the same lines, with the sole difference that everywhere
$\hook{g}{e}$ has to be replaced by $\frac{1}{\theta_g (e,e)}\,\hook{g}{e}$ .}

The representations of $D^{\omega}\!(H)$ can be found in much the same way as
for $D(H)$. It is done by inducing the representations of centralizer
subgroups as follows. Let $\{\,^A\!C\}$ be the set of conjugacy classes of $H$
and introduce a fixed but arbitrary ordering $^A\!C= \{^A\!g_1,\;^A\!g_2,
\ldots,\,^A\!g_k\}$. Let $^A\!N$ be the centralizer of $^A\!g_1$ and
$\{^A\!x_1,\,^A\!x_2,\ldots,\,^A\!x_k\}$ be a set of representatives of the
equivalence classes of $G/^A\!N$, such that $^A\!g_i=\,^A\!x_i\,^A\!g_1\,
^A\!x_i^{-1}$. Choose for convenience $^A\!x_1=e$. Taking different
sets of representatives yield
representations that only differ by an unitary transformation. Consider the
complex vectorspace $V^A_{\alpha}$ spanned by the basis
$\{|^A\!g_j,\,^{\alpha}\!v_i\!>\}_{j=1,\ldots,k}^{i=1,\ldots,
\mbox{\scriptsize dim}\alpha}$. We denote the basis elements of the unitary
irreducible representation $^{\alpha}\!\Gamma$  of $^A\!N$ by
$^{\alpha}\!v_i$. This vectorspace carries a representation $\Pi^A_{\alpha}$
of $\DW$, given by
\be \label{13}
\Pi^A_{\alpha}(\,\hook{g}{x})|\,^A\!g_i,\,^{\alpha}\!v_j \!>=
\delta_{g,x\,^A\!g_ix^{-1}}\;\; \varepsilon_g(x)\;|x\,^A\!g_ix^{-1},
\,^{\alpha}\!\Gamma(\,^A\!x_k^{-1}x\,^A\!x_i)\,^{\alpha}\!v_j \!>,
\ee
with $\,^A\!x_k$  defined through $\,^A\!g_k \equiv x\,^A\!g_ix^{-1}$. The new
ingredient here is the phase $\varepsilon_g$ that is
related to $\theta_g$ by
\be \label{repphase}
\theta_g(x,y)=\frac{\varepsilon_g(x)\varepsilon_{x^{-1}gx}(y)}
{\varepsilon_g(xy)} \;\; ,
\ee
 to make~(\ref{13}) a representation\footnote
{The representation~(\ref{13}) is at variance with (3.3.1) in [dpr1].
We found it unavoidable to include the phase
$\varepsilon$ to obtain true representations of $D^{\omega}(H)$.}.
Associativity of the multipication (\ref{algebra}) guarantees that
$\theta_g$ is a solution of the second `conjugated' cohomology group
\be  \label{2cocycle}
\theta_g(x,y)\theta_g(xy,z)=\theta_g(x,yz)\theta_{x^{-1}gx}(y,z).
\ee
This can be directly verified by substitution of~(\ref{theta}) and
applying~(\ref{pentagon}). Note that equation~(\ref{repphase}) requires that
$\theta_g$ is in fact exact, i.e.\ an element of the trivial class of this
cohomology group. It is not clear to us whether this requirement is met in
general. However, in the examples we have studied so far, $\theta$ indeed
turned out to be exact.

We are now in a position to appreciate the equivalence of the representation
theory of $\DW$ and $D^{\omega\delta\beta}(H)$, and in particular to interpret
the twist freedom~(\ref{F}). At the level of a single
representation~(\ref{13}) the twist~(\ref{twist}, \ref{basistr}) remains
unnoticed. Both the left hand-  and the right hand side of~(\ref{13}) get
multiplied by the phase $\beta(x, x^{-1}gx)/\beta(g,x)$, due to the way
$\varepsilon$ and $\hook{g}{x}$ transform. The transformation property
$\varepsilon_g(x)\, \mapsto \varepsilon_g(x)\;\beta(x,x^{-1}gx)/\beta(g,x)$
under a twist, is inferred from~(\ref{theta}, \ref{twist},
\ref{repphase}).
At the level of the tensorproduct of two representations $\Pi^C_{\gamma} \ot
\Pi^D_{\delta}$, the situation is slightly more involved. From~(\ref{F}), we
see that twisting the algebra is the same as transforming the states in the
tensorproduct $\Pi^C_{\gamma} \ot  \Pi^D_{\delta}$ with $\Pi^C_{\gamma} \ot
\Pi^C_{\delta} (F)$. On the basis vectors, this transformation has the effect
\be
|\,^C\!g_i,\,^{\gamma}\!v_j \!>\!|\,^D\!g_k,\,^{\delta}\!v_l \!> \;\;
\longmapsto   \;\;
\beta (\,^C\!g_i,\,^D\!g_k) \;
|\,^C\!g_i,\,^{\gamma}\!v_j \!>\!|\,^D\!g_k,\,^{\delta}\!v_l \!>,
\ee
i.e.\ these couplings get multiplied by a phase. This effect of the twist is
physically unobservable, since quantumstates are defined up to a
phase.\\[0.5cm]
{\bf Fusion.} By means of Verlinde's formula~\cite{ver0}
\be      \label{verlinde}
N^{AB\gamma}_{\alpha\beta C}=\sum_{D,\delta}\frac{
S^{AD}_{\alpha\delta}S^{BD}_{\beta\delta}
(S^{*})^{CD}_{\gamma\delta}}{S^{eD}_{0\delta}},
\ee
the fusion rules for $\DW$ can be obtained from the modular
$S$ matrix
\be                                \label{fusion}
S^{AB}_{\alpha\beta}=\sum_{\stackrel{\,^A\!g_i\in\,^A\!C\,,^B\!g_j\in\,
^B\!C}{[\,^A\!g_i,\,^B\!g_j]=e}}
\alpha^*(\,^A\!x_i^{-1}\,^B\!g_j\,^A\!x_i)\beta^*(\,^B\!x_j^{-1}\,^A\!g_i\,
^B\!x_j)\,\,\sigma(\,^A\!g_i | \,^B\!g_j),
\ee
with $\alpha(g)\equiv \mbox{Tr}\;^{\alpha}\!\Gamma(g)$ and~\cite{dvvv} the
twist independent phase
$\sigma(g|h) \equiv \varepsilon_g(h)\varepsilon_h(g)$.\\[0.5cm]
{\bf Braiding and Aharonov-Bohm scattering.} The operator ${\cal R}$
that establishes a positively oriented
interchange of two excitations, is associated with the universal
$R$ matrix~(\ref{Rmatrix}). It is defined  as
\[
{\cal R}_{\alpha\beta}^{AB}\equiv
\sigma\circ(\Pi_{\alpha}^A\otimes\Pi_{\beta}^B)(R),
\]
with $\sigma$ the permutation operator~(\ref{permutation}). To be
explicit, the braid operation ${\cal R}$ on the state
$|\,^A\!g_i,\,^{\alpha}\!v_j\!>|\,^B\!g_k,\,^{\beta}\!v_l\!>
\in
\!|\,^A\!C,\,^{\alpha}\!\Gamma>\!\otimes |\,^B\!C,\,^{\beta}\!\Gamma>$
reads
\be                 \label{braidaction}
{\cal R}_{\alpha\beta}^{AB}\;|^A\!g_i,\,^{\alpha}\!v_j\!>\!
|^B\!g_k,\,^{\beta}\!v_l\!> =
\! \varepsilon_{^B\!g_m}(\,^A\!g_i)
|\,^B\!g_m,
\,^{\beta}\!\Gamma(^B\!x_m^{-1\;A}\!g_i^B\!x_k)^{\beta}\!v_j\!> |^A\!g_i,
^{\alpha}\!v_j\!>,
\ee
where $^B\!x_m$ is defined through $^B\!g_m \equiv
\,^A\!g_i\,^B\!g_k\,^A\!g_i^{-1}$.
For a non-trivial 3-cocycle $\omega$, the braid operator ${\cal R}$
does {\em not} satisfy the ordinary -,
but rather the quasi Yang-Baxter equation~\cite{dri1}
\be \label{quasiYBE}
{\cal R}_1  \tilde{\cal R}_2 {\cal R}_1=
\tilde{\cal R}_2   {\cal R}_1  \tilde{\cal R}_2  \; .
\ee
 ${\cal R}_1$ acts on the three-particle states in $(V_1 \ot V_2) \ot V_3$
as ${\cal R}\otimes {\bf 1}$ and $\tilde{\cal R}_2$ as
$\Phi^{-1}\cdot({\bf 1}\otimes {\cal R})\cdot\Phi$, with $\Phi$ the isomorphism
 defined in~(\ref{fi}).
In passing, we mention that the braid operator is  related to
the modular $S$ matrix :
$S=\frac{1}{|H|}\mbox{Tr}\;{\cal R}^2$.

The cross sections of elastic two-particle Aharonov-Bohm scattering
are completely determined by the monodromy matrix ${\cal R}^2$. The
relationship can be expressed as follows~\cite{ver1}
\be \label{Aharonov}
\frac{d\sigma}{d\varphi}=\frac{1}{2\pi k
\sin^2(\varphi/2)}\;\frac{1}{2}\,[1-\mbox{Re}<\!\psi_{in}|{\cal
R}^2|\psi_{in}\!>],
\ee
with $|\psi_{in}\!>$  the incoming two-particle state, and $k$
the relative wave vector.

\section{Examples}
{\bf An abelian example: cocycles on $D(Z_N)$.}
We turn to the example $H \simeq Z_N$, with multiplicative generator $g$,
discussed in detail from the field
theoretic point of view  in section~\ref{model}.
Every element $g^m$
constitutes a conjugacy class and has $Z_{N}$ as its centralizer.
We use $m$ to denote  $g^m$ by $m$, so the group structure is presented
by: $ m_1 \cdot
m_2 = [m_1+m_2] $. Here $[m]$ means $m$ modulo $N$, chosen in the range
$0, \ldots ,N-1$. The $N$ irreducible representations $^l \Gamma$ of this group
are all 1-dimensional, and are given by \be  \label{rep1}
^l \Gamma(m) = e^{ \im \frac{2\pi}{N}lm}.
\ee
Since the group is abelian, the `conjugated' 2-cocycle
condition~(\ref{2cocycle})
coincides with the ordinary 2-cocycle condition.
It is well-established (see for example \cite{mose}) that all even
cohomology groups $H^{even}(Z_{N},U(1))$ are trivial, while for odd
cohomology groups we have $H^{odd}(Z_{N},U(1))\simeq Z_{N}$. Thus $\theta_g \in
H^2(Z_{N},U(1))$ is
trivial and can indeed be written as~(\ref{repphase}). An
explicit realization of  $\omega$ is given by
\be \label{omom}
\omega(m_1,m_2,m_3)= e^{ \im \frac{2\pi p}{N^2}m_1(m_2+m_3-[m_2+m_3])},
\ee
with $p \in [0,\ldots,N-1]$.
{}From~(\ref{omom}) together with~(\ref{theta}, \ref{gamma}), we find
\be   \label{pro}
\theta_{m_1}(m_2,m_3)=\gamma_{m_1}(m_2,m_3)=\omega(m_1,m_2,m_3),
\ee
and consequently $\varepsilon_{m_1}(m_2)=\exp \im \frac{2\pi p}{N^2}m_1m_2$
from~(\ref{repphase}).
Thus the effect of a 3-cocycle $\omega$ on the representations~(\ref{rep1})
is precisely the additional phase occurring in
(\ref{ga+co})
\be \label{ac+co}
\hook{m_1}{m_2}|m,n\!>\,=\delta_{m_1,m}\;\;
e^{ \im \frac{2\pi}{N}m_2(n+p\frac{m}{N})}\;|m,n\!>.
\ee
The 2-cocycle appearing in~(\ref{ray}) is nothing but $\theta$,
see~(\ref{algebra},
\ref{omom}, \ref{pro}).
Implementation of the gauge transformation $\hook{m_1}{m_2}$ on a two particle
state involves the co-multiplication (\ref{algebra}). The extra phase
$\gamma_g$ establishes the anticipated result given in~(\ref{twopart}),
and  the fusion rules
obtained from~(\ref{verlinde}) are the same as~(\ref{csfusion}).
The braid action ${\cal R}$ on the
representations $|m_1,n_1\!>\!\otimes|m_2,n_2\!>$ is given
by~(\ref{braidaction})
\be  \label{11}
{\cal R}\;|m_1,n_1\!>\!|m_2,n_2\!>\,=
e^{ \im \frac{2\pi}{N}m_1(n_2+p\frac{m_2}{N})}\;
|m_2,n_2\!>\!|m_1,n_2\!>.
\ee
We generate the phase~(\ref{csphase}) by repeating this operation,
and obtain the following closed formula for
the Aharonov-Bohm cross-sections~(\ref{Aharonov})
\be \label{crosssections}
\frac{d\sigma}{d\varphi}[(m_1,n_1),(m_2,n_2)]=
\frac{\sin^2 \frac{\pi}{N}(n_1 m_2+ n_2 m_1 +\frac{2p}{N}m_1 m_2)}
{2\pi k \sin^2(\varphi/2)}.
\ee

Note that in this abelian model we find
$\tilde{\cal R}_2={\cal R}_2$ because of the symmetry of $\omega$ in the last
two entries:
$\omega(m_1,m_2,m_3)=\omega(m_1,m_3,m_2)$, see~(\ref{omom}).
This implies
that the  quasi Yang-Baxter equation~(\ref{quasiYBE}) projects down
to the ordinary Yang-Baxter equation ${\cal R}_1 {\cal R}_2 {\cal R}_1=
{\cal R}_2 {\cal R}_1{\cal R}_2$.

In conclusion, the algebaic framework $D^p(Z_N)$ yields exactly the same
result as established in section~\ref{model}. Stated differently,
the introduction
of a Chern-Simons term in our lagrangian~(\ref{action}) is equivalent to
a deformation of  the Hopf algebra $D(Z_N)$ by  a non-trivial 3-cocycle. In the
next
example we study the effect of non-trivial 3-cocycles in the Hopf algebra
$D(H)$ with non-abelian $H$.
It is our conviction that in so doing we are merely studying
the effect of adding a
Chern-Simons term to a discrete non-abelian gauge theory.\\[0.5cm]
{\bf A non-abelian example: cocycles on $D(\bar{D}_N)$.} We now consider the
case of non-abelian $H$, describing non-abelian anyons.
In contrast with the abelian example, the general solution of the 3-cocycle
condition~(\ref{pentagon}) for non-abelian groups is not known to us.
We  have however  numerically evaluated~\footnote{We  thank
Arjan van der Sijs and Alec Maassen van den Brink for
substantial computational aid in the process of numerically solving cocycle
conditions on various finite groups.} the 3-cocycle
condition~(\ref{pentagon}) for all groups up till order 23, and found
among others
\be
H^3(\bar{D}_N,U(1))\simeq Z_{4N}.
\ee
For comparison with the discussion in~\cite{bpm1} we restrict our
considerations to $D(\bar{D}_2)$ with $H^3(\bar{D}_2,U(1))\simeq Z_{8}$. In
other words, $\mu=\mu\bmod 8e^2/4\pi$ and there are 8 inequivalent models. We
have computed $\omega$ from~(\ref{pentagon}) and found that $\theta_g$
defined through equation~(\ref{theta}) is exact.
The ambiguity in the solution of
$\varepsilon$ from~(\ref{repphase}) is due to the 1-dimensional
representations of $\bar{D}_2$, which is the content of the first cohomology
group $H^1(\bar{D}_2,U(1))\simeq Z_2\times Z_2$. In Table~1
we have gathered the phases $\varepsilon_g(h)$ for a particular choice of
1-dimensional representations, and the twist freedom $\beta$.
\begin{table}[t]
\begin{center}
\begin{tabular}{|c||c|c|c|c|c|c|c|c||} \hl
$\varepsilon_g(h)$ & $e$ & $\bar{e}$ & $X_1$ & $\bar{X}_1$ & $X_2$ &
$\bar{X}_2$ & $X_3$ & $\bar{X}_3$     \\ \hl                                \hl
$e$& $1$ & $1$  & $1$ & $1$ & $1$ & $1$ & $1$ & $1$ \\ \hl
$\bar{e}$ & & $\tau^4$ &$1$&$1$&$1$&$1$&$1$&$1$\\ \hl
$X_1$& & & $\tau^{-1} $&$ \tau $&$ \tau^{1/2} $&$ \tau^{1/2} $&
$\tau^{1/2}$ & $\tau^{1/2}$ \\ \hl
$\bar{X}_1$ & & & & $\tau^{-1}$ & $\tau^{1/2}$ & $\tau^{1/2}$ & $\tau^{1/2}$
& $\tau^{1/2}$ \\ \hl
$X_2$ & & & & & $\tau^{-1}$ & $\tau$ & $\tau^{1/2}$ & $\tau^{1/2}$\\ \hl
$\bar{X}_2$ & & & & & & $\tau^{-1}$ & $\tau^{1/2}$& $\tau^{1/2}$ \\ \hl
$X_3$ & & & & & & & $\tau^{-1}$ & $\tau$ \\ \hl
$\bar{X}_3$ & & & & & & & & $\tau^{-1}$\\ \hl
\end{tabular}
\end{center}
\caption{\em The phases $\varepsilon$ for $\bar{D}_2$. We have
chosen the twist freedom such that
$\varepsilon$ is symmetric $\varepsilon_g(h)=\varepsilon_h(g)$, and used the
definition $\tau \equiv \exp \im \pi p/8$ with $p=0,\ldots,8$.}
\end{table}

The fusion algebra again only has half the period of the Chern-Simons
parameter, i.e.\ there are only 4 different sets of fusion rules. As in the
abelian example, we have to turn to the braid properties of the exciations to
distinguish all 8 different Chern-Simons models.       In
\cite{bpm1}, we discussed the fusion rules in absence of a Chern-Simons term.
Here, we only gather some salient fusion rules in Table~2 for the
four inequivalent Chern-Simons models.
\begin{table}[h]
\begin{center}
\begin{tabular}{c|cccc}
fusion rule               &    p=0      &   p=1    &     p=2   &     p=3    \\
\hline
$\bar{1}\times\sigma^{\pm}_a$&$\sigma^{\pm}_a$&$\tau^{\mp}_a$
&$\sigma^{\mp}_a$&$\tau^{\pm}_a$ \\
$\bar{1}\times\tau^{\pm}_a$&$\tau^{\mp}_a$&$\sigma^{\mp}_a$
&$\tau^{\pm}_a$&$\sigma^{\pm}_a$ \\
$\sigma^{\pm}_a\times\sigma^{\pm}_a$ & $1+\bar{1}+J_a+\bar{J}_a$&
$1+J_a+\bar{\phi}$& $1+J_a+\sum_{c\neq a}\bar{J}_c$&$1+J_a+\bar{\phi}$ \\
$\sigma^{\pm}_a\times\sigma^{\pm}_b$ & $\sigma^+_c+\sigma^-_c$&
$\sigma^+_c+\sigma^-_c$&$\sigma^+_c+\sigma^-_c$&$\sigma^+_c+\sigma^-_c$ \\
$\sigma^{\pm}_a\times\tau^{\pm}_a$ & $\phi+\bar{\phi}$&
$\phi+\sum_{c\neq a}\bar{J}_c$&$\phi+\bar{\phi}$&$\phi+\sum_{c\neq
a}\bar{J}_c$
\end{tabular}
\end{center}
\caption{\em Some  fusion rules of a discrete $\bar{D}_2$ gauge theory
for different values of the Chern-Simons parameter $p$. The fusion rules are
periodic in $p$ with period $4$.}
\end{table} We use the same notation as in
\cite{bpm1}. The fusion rules~(\ref{verlinde}) only involve $\varepsilon_g(h)$
in the twist independent combination $\sigma(g|h)=\varep_g(h)\varep_h(g)$
where $g$ and $h$ commute. So to acquire them we do not
need the complete solution for $\varepsilon$  given in
Table~1. In fact, if we are only interested in the fusion rules, a more
economical way of calculating them is along the lines described in~\cite{dvvv},
where the minimal set of defining equations for $\sigma(g|h)$ is derived.

The Aharonov-Bohm scattering of particles with fluxes that do not commute
involves all $\varepsilon$'s.   We illustrate this for the example in which a
$\sigma_1^+$ excitation is scattered off  a $\sigma_2^-$ excitation.
The $\sigma_i^{\pm}$ excitation is associated with the conjugacy class
$^{X_i}C = \{X_i , \bar{X}_i \}$. We choose the representatives as
$^{X_1}x_1=\!^{X_2}x_2=e$,   $^{X_1}x_2=X_2$ and $^{X_2}x_2=X_3$. Now suppose
that this 2 particle system is described by the four component quantumstate
\[
|\psi_{in}\!>=(\cos\theta |X_1,^0\!v \!> + \sin\theta |\bar{X}_1,^0\!v \!>)
            (\cos\theta' |X_2,^2\!v\!> + \sin\theta'|\bar{X}_2,^2\!v\!>).
\]
The monodromy operation ${\cal R}^2$~(\ref{braidaction})  on this state yields
\beas
{\cal R}^2 |\psi_{in}\!>&=&
    \varep_{\bar{X}_1}(\bar{X}_2)\varep_{\bar{X}_2}(X_1) \cos\theta
    \cos\theta' |\bar{X}_1,^0\!v \!>\!|\bar{X}_2,-^2\!v\!> +\\
                        & &
    \varep_{\bar{X}_1}(X_2)\varep_{X_2}(X_1) \cos\theta
    \sin\theta' |\bar{X}_1,^0\!v \!>\!|X_2,-^2\!v\!> +\\
                        & &
    \varep_{X_1}(\bar{X}_2)\varep_{\bar{X}_2}(\bar{X}_1) \sin\theta
    \cos\theta' |X_1,^0\!v \!>\!|\bar{X}_2,-^2\!v\!> +\\
                        & &
    \varep_{X_1}(X_2)\varep_{X_2}(\bar{X}_1) \sin\theta
    \sin\theta' |X_1,^0\!v \!>\!|X_2,-^2\!v\!>.
\eeas
If we take the inproduct $<\! \psi_{in}|{\cal R}^2 |\psi_{in}\!>$, substitute
the $\varep$'s of Table~1 and finally take the real part, then we obtain
the following cross-section
\[
\frac{d\sigma}{d\varphi}(\sigma_1^+,\sigma^-_2)=\frac{1}{2\pi k
\sin^2(\varphi/2)}\;\,\frac{1}{2}\;\,[1 + \cos\frac{\pi p}{8}
\sin 2\theta  \sin 2\theta'].
\]

\section{Concluding remarks}
We have found that the  algebraic framework underlying discrete $H$ gauge
theories with Chern-Simons term is the quasi Hopf algebra $\DW$, i.e.\ the
Chern-Simons term introduces a 3-cocycle $\omega$ on the Hopf algebra
$D(H)$.

The framework obviously extends to some of the continous gauge groups when we
view them as limiting cases of discrete gauge groups. In particular the gauge
group $U(1)\times Z_2$ that is relevant for the description of the Alice
string corresponds to $D_{\infty}$. It would be interesting to
review the discussions on the Alice string in our framework.

Our observations may be relevant  in the setting of the fractional
quantum Hall effect. In that respect it would be
useful to know how to compute the anyonic wavefunctions using the connection
with conformal blocks of the appropriate orbifold. This is subject of further
scrutiny.

{}From the point of view of conformal field theory it is of interest to
mention that the fusion rules of $D^{\omega}(\bar{D}_2)$ for $p=1$ {\em
coincide} with the level 1 $SU(2)/D_2$-orbifold~\cite{dvvv} after modding out
the appropriate $Z_2$ generated by $\bar{1}$ (see Table~2). Apparently, the
algebraic
structure of such non-holomorphic orbifolds is still determined by the
`holomorphic' Hopf algebra, be it deformed by a non-trivial 3-cocycle.
To our knowledge, this has not been noticed before.

\subsection*{Acknowledgments}
We wish to thank Arjan van der Sijs and Alec Maassen van den Brink for usefull
discussions. This work has been partially supported by
the Dutch Science Organization FOM/NWO.

\end{document}